\documentstyle[preprint,eqsecnum,aps]{revtex}
\tightenlines
\begin{document}

\title{Entropies  of general nonextreme stationary
axisymmetric black hole: statistical mechanics  and thermodynamics
 $^*$\footnotetext[1]{
This work was partially supported by the National Nature Science
Foundation of China under Grant No. 19975018, and Nature Science
Foundation of Hunan Province}}
\author{Jiliang Jing $^{a b }$  \ \ \ \ Mu-Lin Yan $^{a}$}
\address{a)Department of Astronomy and Applied Physics, University of Science
and Technology of China,\\ Hefei, Anhui 230026, P. R. China; \\ b)
Physics Department and Institute of Physics, Hunan Normal
University,\\ Changsha, Hunan 410081, P. R. China}

\maketitle
\begin{abstract}
Starting from metric of  general nonextreme stationary
axisymmetric black hole in four-dimensional spacetime, both
statistical-mechanical and thermodynamical entropies are studied.
First, by means of the ``brick wall" model in which the Dirichlet
condition is replaced by a scattering ansatz for the field
functions at the horizon and with the Pauli-Villars regularization
scheme, an expression for the statistical-mechanical entropy
arising from the nonminimally coupled scalar fields is obtained.
Then, by using the conical singularity method Mann and
Solodukhin's result for the Kerr-Newman black hole (Phys. Rev.
D54, 3932(1996)) is extended to the general stationary black hole
and nonminimally coupled scalar fields.  We last show by comparing
the two results that the statistical-mechanical entropy and
one-loop correction to the thermodynamical entropy are equivalent
for coupling $\xi\leq 0$. After renormalization, a relation
between the two entropies is given.

\end{abstract}

\newpage

\section{INTRODUCTION}
\label{sec:intro} \vspace*{0.5cm}

There are two definitions of entropy of fields on the black hole
exterior. One definition is based on the covariant Euclidean
formulation, another one is canonical. The thermodynamical entropy
of a black hole is related to the covariant Euclidean free energy
$F^E[g, \beta]=\beta^{-1}W[g, \beta]$\cite{Frolov98}, where
$\beta$ is the inverse temperature. Function $W[g, \beta]$ is
given on Euclidean manifolds with the period $\beta$ in Euclidean
time $\tau$. An alternative approach to calculate $F^E$ is the
conical singularities method \cite{Solodukhin951}-\cite{Mann96}.
The statistical-mechanical entropy can be derived from the
canonical formulation \cite{Frolov98}. The corresponding free
energy, $F^C$, can be defined in term of the one-particle
spectrum. One of the ways to calculate $F^C$ is ``brick wall"
model (BWM) proposed by 't Hooft \cite{Hooft85}. In the model, in
order to eliminate divergence which appears due to the infinite
growth of the density of states close to the horizon, 't Hooft
introduces a ``brick wall" cutoff: a fixed boundary near the event
horizon within the quantum field does not propagate and the
Dirichlet boundary condition is imposed on the boundary. In order
to get the correct $\xi$ (coupling constant) dependence for the
statistical-mechanical entropy and obtained the correspondence
with the conical singularity method, Solodukhin
\cite{Solodukhin97} introduced a scattering ansatz, $i.e.,$ the
field function near the event horizon that describes scattering by
the hole with some nontrivial change of phase.

For ordinary thermodynamical systems it is well known that the
thermodynamical and the statistical-mechanical entropies are
exactly the same. For the black holes, the study of the two
entropies has attracted much attention recently\cite{Frolov98}
\cite{Hooft85}-\cite{Fursaev97}. The study was motivated by
attempts to explain the entropy of black holes as the
statistical-mechanical entropy of quantum fields propagating near
the event horizon. In Ref.\cite{Solodukhin96} Solodukhin
demonstrated in the two-dimensional example that the
thermodynamical entropy of a black hole coincides with its
statistical-mechanical one. In Ref. \cite{Solodukhin97} Solodukhin
calculated the statistical-mechanical entropy for a scalar filed
with nonminimally coupling in the four-dimensional static black
hole spacetime and found that for $\xi\leq 0 $ the result agrees
with one-loop correction to the thermodynamical
entropy\cite{Solodukhin953}. Frolov and Fursaev\cite{Frolov98}
reviewed studies of the relation between the thermodynamic entropy
and the statistical-mechanical entropy of black holes. They showed
that the covariant Euclidean free energy $F^E$ and the canonical
free energy $F^C$ are equivalent when ones use the ultraviolet
regularization \cite{Frolov98} for the general static black holes.

Mann and Solodukhin expected in Ref. \cite{Mann96} that study of
the two entropies for the stationary axisymmetric black hole
should provide us with better understanding of the relationship
between the different entropies. It is general opinion that the
study of the quantum entropies for the stationary axisymmetric
black hole in four-dimensional spacetime is an interesting topic
and should be investigated deeply. Therefore, much effort has also
been contributed the research in the last years\cite{Mann96}
\cite{Cognola98}\cite{Lee96} \cite{Lee961}. Mann and Solodukhin
\cite{Mann96} studied thermodynamical entropy of the Kerr-Newman
black hole  due to a minimally coupled scalar field by using the
conical singularities method. Cognola \cite{Cognola98}, through
the Euclidean path integral and using a heat kernel and
$\zeta$-function regularization scheme, studied the one-loop
contribution to the entropy for a scalar field in the Kerr black
hole (he pointed that the result is valid also for the Kerr-Newman
black hole). Unfortunately, the result is in contrast with the
corresponding one obtained by the conical singularities
\cite{Mann96}.

Recently, we \cite{Jing99} studied the statistical-mechanical
entropy by using the BWM and with the Pauli-Villars regularization
scheme. We showed that one-loop correction to the  thermodynamical
entropy and the statistical-mechanical entropy due to the scalar
quantum fields are equivalent for the Kerr-Newman and the
Einstein-Maxwell dilaton-axion black holes. However, at the moment
the relation between two entropies for general nonextreme
stationary axisymmetric black hole still remains open. The aim of
this paper is to settle the question for nonextreme case.

The  paper is organized  as follows: In Sec. II, by using the BWM
\cite{Solodukhin97} we deduce  a formula of statistical entropy
arising from nonminimally coupled scalar fields  in the general
nonextreme stationary axisymmetric  black hole spacetime. In sec.
III, by means of the conical singularities method we extend Mann
and Solodukhin's result  in Ref. \cite{Mann96} to the nonminimally
coupled scalar fields and the general stationary axisymmetric
black hole. Two results are then compared and a summary is
presented in last section. Some calculations are given in
Appendices.

\section{statistical-mechanical entropy  of general \\
nonextreme stationary axisymmetric black holes} \vspace*{0.5cm}

The metric for the general stationary axisymmetric black hole in
Boyer-Lindquist coordinates can be expressed as
\begin{equation}
ds^2=g_{tt}dt^2+g_{rr}dr^2+g_{t \varphi}dtd\varphi+g_{\theta
\theta}d\theta ^2+g_{\varphi \varphi}d\varphi ^2,
 \label{metric0} \end{equation}
where $g_{tt}$, $g_{rr}$, $g_{t\varphi}$, $g_{\theta \theta}$ and
$g_{\varphi \varphi}$ are functions of the coordinates $r$ and
$\theta$ only. For the nonextreme case  we have relations (see
\cite{Jing99} for details)
\begin{eqnarray}
\left(g_{tt}-\frac{g_{t\varphi}^2}{g_{\varphi\varphi}}
\right)&=&G_1(r,\theta)(r-r_H), \label{asum}\\
g^{rr}&=&G_2(r,\theta)(r-r_H),\label{asum1}\\
g_{rr}\left(g_{tt}-\frac{g_{t\varphi}^2}{g_{\varphi\varphi}}
\right)&=&\frac{G_1(r, \theta)}{G_2(r, \theta)} \equiv
-f(r,\theta), \label{relation}
\end{eqnarray}
where $r_H$ represents event horizon, $G_1(r,\theta)$,
$G_2(r,\theta)$ and $f(r, \theta)$ are regular functions on and
outside the event horizon.

We now try to find a general statistical-mechanical entropy
expression for the space-time (\ref{metric0}).  We first seek
quantization condition by using the motion equation of the scalar
field and introducing the scattering boundary condition,  and then
use the condition to calculate  free energy. The
statistical-mechanical entropy is obtained by the variation of the
corresponding free  energy with respect to the inverse Hawking
temperature.

Using the a WKB approximation with
\begin{equation}
\phi =exp[-iEt+im\phi+iW(r,\theta)]=
exp[-iEt+im\phi]\psi(r, \theta), \label{phi}
\end{equation}
and substituting  the metric (\ref{metric0}) into  the motion
equation of the  scalar field with mass $\mu$ and arbitrary
coupled to the scalar curvature $R$
\begin{equation}
\frac{1}{\sqrt{-g}}\partial _\mu(\sqrt{-g}g^{\mu\nu}\partial _\nu
\phi)-(\mu ^2+\xi R)\phi=0, \label{kg}
\end{equation}
We know from discussion in Ref.\cite{Jing99} that the function
$W(r,\theta)$ can be expressed as
 \begin{equation}
W(r, \theta)=\pm \int^r
\sqrt{\frac{-g_{rr}g_{\varphi\varphi}}{g_{tt}g_{\varphi\varphi}-
g_{t\varphi }^2}} K(r, \theta) dr+c(\theta),
\end{equation}
where
 \begin{equation}
K(r, \theta)=\sqrt{(E-\Omega m)^2+\left(g_{tt}-
\frac{g_{t\varphi}^2}{g_{\varphi\varphi}}
\right)\left(\frac{m^2}{g_{\varphi\varphi}}+\frac{p_\theta^2}
{g_{\theta\theta}}+M^2(r, \theta)\right)}.\label{KK}
\end{equation}
 Then, the function $\psi(r, \theta)$ in (\ref{phi}) is given by
\begin{eqnarray}
\psi(r,
\theta)&=&exp\left[i\int^r{\sqrt{\frac{-g_{rr}g_{\varphi\varphi}}{g_{tt}
g_{\varphi\varphi}-g_{t\varphi}^2}}K(r, \theta)dr} \right]
\nonumber\\& &+A
exp\left[-i\int^r{\sqrt{\frac{-g_{rr}g_{\varphi\varphi}}{g_{tt}
g_{\varphi\varphi}-g_{t\varphi}^2}}K(r, \theta)dr}
\right],\label{ff}
\end{eqnarray}
where the constant $A$ is to be determined with the boundary
conditions. In Eq.  (\ref{ff}) the amplitude is a slowly varying
function and is omitted in writing. At the boundary $\Sigma_{h}$
staying at a small distance $h$ from the event horizon $\Sigma$,
Solodukhin's  scattering condition \cite{Solodukhin97} is shown by
 \begin{equation}
(n^\mu \partial _\mu \phi -\xi k \phi)|_{\Sigma_h}=0,\label{b1}
\end{equation}
where $n^\mu$ is vector normal to $\Sigma_h$ and $k$ is extrinsic
curvature of $\Sigma_h$. For the stationary axisymmetric black
hole (\ref{metric0}), after setting
\begin{equation}
n_\mu=(0,\sqrt{g_{rr}},0,0),
\end{equation}
we find the  extrinsic curvature  can be written as\cite{Jing96}
\begin{eqnarray}
k&=&\frac{\sqrt{g_{rr}}}{2g_{rr}g_{\theta
\theta}(g_{tt}g_{\varphi\varphi}-g_{t\varphi}^2)}\left[
(g_{tt}g_{\varphi \varphi}-g_{t\varphi}^2)\frac{\partial
g_{\theta\theta}} {\partial r}\right.\nonumber\\&
&+\left.g_{tt}g_{\theta \theta }\frac{\partial g_{\varphi \varphi
}}{\partial r}-2g_{t\varphi }g_{\theta \theta}\frac{\partial
g_{t\varphi }}{\partial r}+g_{\varphi \varphi}g_{\theta
\theta}\frac{\partial g_{tt}}{\partial r}\right], \label{k1}
\end{eqnarray}
on the boundary $\Sigma_h$ Eq.  (\ref{k1}) reduces to
 \begin{equation}
k(\Sigma_h)\approx \left[\frac{1}{2\sqrt{g_{rr}}\left(g_{tt}
-\frac{g_{t\varphi}^2}{g_{\varphi
\varphi}}\right)}\frac{\partial }{\partial r}
\left(g_{tt}-\frac{g_{t\varphi}^2}{g_{\varphi \varphi}}
\right)\right]_{\Sigma_h}. \label{k2}
\end{equation}
From Eqs.  (\ref{ff}), (\ref{k2}) and (\ref{b1}) we have
 \begin{equation}
\left[\sqrt{\frac{-\left(g_{tt}-\frac{g_{t\varphi}^2}{g_{\varphi
\varphi}}\right)}{g_{rr}}}\frac{\partial \psi(r, \theta)}{\partial
r}-\xi^*\psi(r, \theta) \right]_{\Sigma_h}=0, \label{b2}
\end{equation}
where $\xi^*\equiv\frac{2\pi \xi}{\beta _H}$ and $\beta _H$ is the
Hawking inverse temperature. Making use of Eq. (\ref{b2}) and
another boundary condition $\phi=0,$ for $ r=r_E$ ($r_E<r_{VLS}$,
where $r_{VLS}$ is position of the velocity of the light surface
\cite{Lee96}\cite{Lee961}), after discussion as Solodukhin
\cite{Solodukhin97} we find following quantum condition
 \begin{equation}
2\int^{r_E}_{r_H+h}\sqrt{\frac{-g_{rr}g_{\varphi
\varphi}}{g_{tt}g_{\varphi
\varphi}-g_{t\varphi}^2}} K(r,\theta) dr=\nu \eta (K)+\pi \nu +2\pi
n(E, m, p_{\theta}, \theta), \label{n}
\end{equation}
with
\begin{equation}
\eta(K)=tan^{-1} \left(\frac{2K(h)\xi^*}{K(h)^2-\xi^{*2}}\right),
\ \ \ \ \ \ \nu=2\xi.\label{eta}
\end{equation}
We can separate $n(E, m, p_{\theta}, \theta)$ into  two parts
\begin{equation}
n(E, m, p_{\theta}, \theta)=n_0(E, m, p_{\theta}, \theta) +n_1(E,
m, p_{\theta}, \theta).
\end{equation}
From Eqs. (\ref{n}) and (\ref{eta}) we find  that $ n_0(E, m,
p_{\theta}, \theta)$ and $n_1(E, m, p_{\theta}, \theta)$ can be
respectively expressed as
\begin{equation} n_0(E, m,
p_{\theta})=\frac{1}{\pi} \int d\theta
\int^{r_E}_{r_H+h}\sqrt{\frac{-g_{rr}g_{\varphi
\varphi}}{g_{tt}g_{\varphi \varphi}-g_{t\varphi}^2}}K(r,\theta)dr,
\end{equation}
\begin{equation}
n_1(E, m, p_{\theta})=\frac{-\nu}{2\pi}\int \left[tan^{-1}
\left(\frac{2K(h)\xi^*}{K(h)^2-\xi^{*2}} \right)+\pi
\right]_{\Sigma_h} d\theta.
\end{equation}
It is shown that $n_0(E, m, p_{\theta}, \theta)$ represents the
original 't Hooft particles\cite{Solodukhin97}\cite{Jing99},  and
$n_1(E, m, p_{\theta}, \theta)$ the scattering
particles\cite{Solodukhin97}.

For an ensemble of states of the scalar field, the free  energy
can be expressed as
\begin{eqnarray}
\beta F&=& \int dm \int dp_{\theta}\int dn(E, m, p_{\theta})ln \left[
1-e^{-\beta (E-\Omega_0 m)}\right] \nonumber \\
&=&\int dm \int dp_{\theta}\int dn(E+\Omega_0 m, m, p_{\theta})ln \left(
1-e^{-\beta E}\right) \nonumber \\
&=&-\beta \int dm \int dp_{\theta}\int
\frac{n_0(E+\Omega_0 m, m, p_{\theta})+n_1(E+\Omega_0 m, m, p_{\theta})}
{e^{\beta E}-1} dE \nonumber \\
&=&-\beta \int \frac{n_0(E)+n_1(E)}{e^{\beta E}-1} dE \nonumber \\
&=&\beta F_0 + \beta F_1, \label{f1}
\end{eqnarray}
with
\begin{eqnarray}
n_0(E)&=&\int dm \int dp_{\theta}\int n_0(E+\Omega_0 m, m,
p_{\theta}) \nonumber \\ &=&\frac{1}{3\pi}\int d\theta
\int^{r_E}_{r_H+h}\frac{\sqrt{g_4} \left[E^2
+\left(g_{tt}-\frac{g_{t\varphi}^2}{g_{\varphi \varphi}}
\right)\left(1+\frac{g_{\varphi
\varphi^2(\Omega-\Omega_0)^2}}{g_{tt}g_{\varphi \varphi}-g_{t
\varphi}^2} \right)M^2(r,\theta)\right]^{\frac{3}{2}}}
{\left[\left(g_{tt}-\frac{g_{t\varphi }^2}{g_{\varphi
\varphi}}\right)\left(1+\frac{g_{\varphi
\varphi^2(\Omega-\Omega_0)^2}}{g_{tt}g_{\varphi \varphi}-g_{t
\varphi}^2} \right)\right]^2} ,\nonumber \\ \label{n01}
\end{eqnarray}
\begin{eqnarray}
n_1(E)&=&\int dm \int dp_{\theta}\int n_1(E+\Omega_0 m, m,
p_{\theta}) \nonumber \\ &\approx&\frac{\nu}{\pi}\int d\theta
\left\{\frac{ \sqrt{g_{\theta \theta}g_{\varphi \varphi}}\left[
\xi^*\bar{K}-\xi^{*2}tan^{-1} \left(\frac{\bar{K}}{\xi^*}
\right)+\frac{\bar{K}^2}{2}tan^{-1} \left(\frac{2\xi^*\bar{K}}
{\bar{K}^2-\xi^{*2}}\right)\right]}{\left(g_{tt}-
\frac{g_{t\varphi}^2}{g_{\varphi
\varphi}}\right)\left(1+\frac{g_{\varphi
\varphi^2(\Omega-\Omega_0)^2}}{g_{tt}g_{\varphi \varphi}-g_{t
\varphi}^2} \right)} \right\}_{\Sigma_{h}}. \nonumber
\\ \label{n11}
\end{eqnarray}

Taking the integration $r$ of Eq. (\ref{n01}) for
$\Omega_0=\Omega_H$ and focusing only on the divergent
contribution at horizon, we find
\begin{eqnarray}
n_0(E)&=&-\frac{1}{2\pi}\int d\theta \left \{\sqrt{g_{\theta
\theta}g_{\varphi \varphi}}\left[\frac{2}{3}\left(\frac{E
\beta_H}{4\pi} \right)^3C(r,\theta)+M^2(r,\theta) \left(\frac{E
\beta_H}{4\pi}\right)\right]\right. \nonumber \\& &\times\left. ln
\left(\frac{E^2}{E^2_{min}}\right) \right\}_{r_H}
-\frac{1}{3\pi}\left(\frac{\beta_H}{4\pi}\right)\int d\theta
\left\{\sqrt{g_{\theta \theta}g_{\varphi\varphi}}M^2(r,\theta)
\left(E-\frac{E^3}{E^2_{min}}\right)\right\}_{r_H} , \nonumber\\
\label{n0}\\ n_1(E)&=&\frac{\nu}{\pi}\int d\theta
\left\{\frac{(\xi^{*}M(r,\theta ))^2\sqrt{g_{\theta
\theta}g_{\varphi \varphi}}}{E^2_{min}}\left[
tan^{-1}\Big(\frac{\bar{K}}{\xi^*}
\Big)-\frac{\bar{K}}{\xi^*}\right.\right.\nonumber \\ & &
-\left.\left.\frac{\bar{K}^2}{2\xi^{*2}}tan^{-1}
\Big(\frac{2\xi^*\bar{K}}
{\bar{K}^2-\xi^{*2}}\Big)\right]\right\}_{r_{H}}, \nonumber \\
\label{n1}
\end{eqnarray}
where
\begin{eqnarray}
C(r,\theta)&=&\frac{\partial ^2g^{rr}}{\partial
r^2}+\frac{3}{2}\frac{\partial g^{rr}}{\partial r}\frac{\partial
\ln f}{\partial r}-\frac{1}{2}\frac{\partial g^{rr} }{\partial r
}\left(\frac{1}{g_{\theta \theta}}\frac{\partial g_{\theta
\theta}}{\partial r}\right.\nonumber\\& &+\left.
\frac{1}{g_{\varphi \varphi}}\frac{\partial g_{\varphi
\varphi}}{\partial r}\right)-\frac{2g_{\varphi
\varphi}}{f}\left[\frac{\partial}{\partial r}\left(\frac{g_{t
\varphi}}{g_{\varphi \varphi}}\right)\right]^2,\nonumber \\
\label{c1}
\\
 E^2_{min}&=&-M^2(r_H, \theta)\left(g_{tt}-\frac{g_{t\varphi^2}}{g_{\varphi
\varphi}}\right)_{\Sigma_h}, \nonumber \\
\bar{K}^2&=&E^2+\left(g_{tt}-\frac{g_{t\varphi}^2}{g_{\varphi
\varphi}}\right)_{\Sigma_h}M^2(r_H, \theta).
\end{eqnarray}

Let use the Pauli-Villars regularization scheme \cite{Demers95} by
introducing five regulator fields $\{\phi_i, i=1,...,5\}$ of
different statistics with masses $\{\mu_i,i=1,...,5\}$ dependent
on the UV cutoff \cite{Demers95} and with the same nonminimal
coupling $\{\xi_i=\xi, i=0,...,5\}$. If we rewrite the original
scalar field $\phi=\phi_0$ and $\mu=\mu_0$, then these fields
satisfy $\Sigma^5_{i=0}\triangle _i=0$ and
$\Sigma^5_{i=0}\triangle _i \mu^2_i=0$, where
$\triangle_0=\triangle_3=\triangle_4=+1$ for the commuting fields
and $\triangle_1=\triangle_2=\triangle_5=-1$ for the anticommuting
fields. Since each of the fields makes a contribution to the free
energy of Eq. (\ref{f1}), and the total free energy becomes
\begin{eqnarray}
\beta \bar{F}=\sum^5_{i=0}\triangle _i \beta
F_i= \beta \bar{F_0}+\beta \bar{F_1}.\label{f2}
\end{eqnarray}
Substituting Eqs. (\ref{f1}), (\ref{n0}), and (\ref{n1}) into
(\ref{f2}) and integrating over E we find
\begin{eqnarray}
\bar{F_0}&=&-\frac{1}{48}\frac{\beta_H}{\beta ^2}\int
d\theta\left\{\sqrt{g_{\theta \theta}g_{\varphi
\varphi}}\right\}_{r_H} \sum^5_{i=0}\triangle _i
M_i^2(r_H,\theta)lnM_i^2(r_H, \theta) \nonumber \\ & &
-\frac{1}{2880}\frac{\beta_H^3}{\beta^4} \int
d\theta\left\{\sqrt{g_{\theta \theta}g_{\varphi \varphi}}
\left[\frac{\partial ^2g^{rr}}{\partial r^2}+
\frac{3}{2}\frac{\partial g^{rr}}{\partial r}\frac{\partial \ln
f}{\partial r}-\frac{1}{2}\frac{\partial g^{rr}}{\partial
r}\left(\frac{1}{g_{\theta \theta}}\frac{\partial g_{\theta
\theta}}{\partial r}\right.\right.\right.\nonumber\\& &
+\left.\left.\left. \frac{1}{g_{\varphi \varphi}}\frac{\partial
g_{\varphi \varphi}}{\partial r}\right) -\frac{2g_{\varphi
\varphi}}{f}\left[\frac{\partial}{\partial r}\left(\frac{g_{t
\varphi}}{g_{\varphi \varphi}}\right)\right]^2
\right]\right\}_{r_H}\sum^5_{i=0}\triangle _ilnM_i^2(r_H,\theta)
\label{f-0}
\end{eqnarray}
\begin{eqnarray}
\bar{F_1}&=&-\frac{|\xi|}{4\beta }\int
d\theta\left\{\sqrt{g_{\theta \theta}g_{\varphi \varphi}}\right\}_{r_H}
\sum^5_{i=0}\triangle _i M_i^2(r_H,\theta)lnM_i^2(r_H,
\theta),\label{f-1}
\end{eqnarray}
where $M_i^2(r_H, \theta)=\mu_i^2-\left(\frac{1}{6}-\xi\right)R$.
The total entropy at the Hawking temperature
$\frac{1}{\beta}=\frac{1}{\beta_H}$ is given by
\begin{eqnarray}
S^{SM}&=&\left[\beta^2 \frac{\partial \bar{F}}{\partial
\beta}\right]_{\beta=\beta_H}=\left[\beta^2 \frac{\partial
(\bar{F_0}+\bar{F_1})}{\partial
\beta}\right]_{\beta=\beta_H}\nonumber \\
&=&\frac{1}{4}\left(\frac{1}{6}+|\xi|\right)\int
d\theta\left\{\sqrt{g_{\theta \theta}g_{\varphi
\varphi}}\right\}_{r_H} \sum^5_{i=0}\triangle _i
M_i^2(r_H,\theta)lnM_i^2(r_H, \theta) \nonumber
\\& &+\frac{1}{720}\int d\theta \left\{\sqrt{g_{\theta
\theta}g_{\varphi \varphi}} \left[\frac{\partial
^2g^{rr}}{\partial r^2}+ \frac{3}{2}\frac{\partial
g^{rr}}{\partial r}\frac{\partial \ln f}{\partial
r}-\frac{1}{2}\frac{\partial g^{rr}}{\partial
r}\left(\frac{1}{g_{\theta \theta}}\frac{\partial g_{\theta
\theta}}{\partial r}\right.\right.\right.\nonumber \\ & &+
\left.\left. \left. \frac{1}{g_{\varphi \varphi}}\frac{\partial
g_{\varphi \varphi}}{\partial r}\right)-\frac{2g_{\varphi
\varphi}}{f}\left[\frac{\partial}{\partial r}\left(\frac{g_{t
\varphi}}{g_{\varphi
\varphi}}\right)\right]^2\right]\right\}_{r_H}
 \sum^5_{i=0}\triangle _ilnM_i^2(r_H,
\theta). \nonumber \\ \label{SM}
\end{eqnarray}
Using  the assumption that the scalar curvature at the horizon is
much smaller than each $\mu_i$ and noting that the area of the
event horizon can be expressed as $A_{\Sigma}= \int d\varphi \int
d\theta\left\{\sqrt{g_{\theta \theta}g_{\varphi
\varphi}}\right\}_{r_H}$, we obtain following  expression
\begin{eqnarray}
&&S^{SM}=\nonumber \\&&
\frac{A_{\Sigma}}{48\pi}\left(1+6|\xi|\right)\sum^5_{i=0}
\triangle _i \mu_i^2ln\mu_i^2+\left\{-\frac{1}{8\pi}
\left(\frac{1}{6}+|\xi|\right) \left(\frac{1}{6}-\xi\right)\int
d\varphi d\theta\left(R\sqrt{g_{\theta \theta}g_{\varphi
\varphi}}\right)_{r_H} \right.\nonumber \\ & & +\left. \int
d\varphi d\theta\left[\frac{\sqrt{g_{\theta \theta}g_{\varphi
\varphi}}}{1440\pi } \left( \frac{\partial ^2g^{rr}}{\partial
r^2}+ \frac{3}{2}\frac{\partial g^{rr}}{\partial r}\frac{\partial
\ln f}{\partial r}-\frac{1}{2}\frac{\partial g^{rr}}{\partial
r}\left(\frac{1} {g_{\theta \theta}}\frac{\partial g_{\theta
\theta}}{\partial r}+ \frac{1}{g_{\varphi \varphi}}\frac{\partial
g_{\varphi \varphi}}{\partial r}\right)\right. \right. \right.
\nonumber \\ & & \left. \left. \left. -\frac{2g_{\varphi
\varphi}}{f}\left[\frac{\partial}{\partial r}\left(\frac{g_{t
\varphi}}{g_{\varphi
\varphi}}\right)\right]^2\right)\right]_{r_H}\right\}
\sum^5_{i=0}\triangle _iln\mu_i^2.  \label{smuu}
\end{eqnarray}
If we set $\xi=0$  we known that Eq. (\ref{smuu}) gives result of
the Ref.\cite{Jing99}.

The main aim of this paper is to seek the relation between the
statistical-mechanical and the thermodynamical entropies. In order
to do that, we will cast result (\ref{smuu}) into another form. In
appendix A we proved
\begin{eqnarray}
 & &\left\{ \frac{\partial ^2g^{rr}}{\partial r^2}+
\frac{3}{2}\frac{\partial g^{rr}}{\partial r}\frac{\partial \ln
f}{\partial r}-\frac{1}{2}\frac{\partial g^{rr}}{\partial
r}\left(\frac{1} {g_{\theta \theta}}\frac{\partial g_{\theta
\theta}}{\partial r}+ \frac{1}{g_{\varphi \varphi}}\frac{\partial
g_{\varphi \varphi}}{\partial r}\right) -\frac{2g_{\varphi
\varphi}}{f}\left[\frac{\partial}{\partial r}\left(\frac{g_{t
\varphi}}{g_{\varphi
\varphi}}\right)\right]^2\right\}_{r_H}\nonumber \\ & &=
R_{aa}(r_H,\theta)-2R_{abab}(r_H,\theta), \label{raa}
\end{eqnarray}
where $R_{aa}=\sum^2_{a=1}R_{\mu \nu} n^{\mu}_{a}n^{\nu}_{a}, \ $
$R_{abab}=\sum^2_{a, b=1}R_{\mu \nu \lambda \rho}
n^{\mu}_{a}n^{\nu}_{b} n^{\lambda}_{a}n^{\rho}_{b}$ are  the
projections of the curvature onto the subspace normal to the
horizon surface and $\{n^{\mu}_a,\  a=1,2\}$ (see Eq.  (\ref{a11})
in Appendix A) are a pair of vectors orthogonal to the event
horizon $\Sigma$. From (\ref{raa}) we know that Eq.  (\ref{smuu})
can be rewritten as
\begin{eqnarray}
S^{SM}&=&\frac{A_{\Sigma}}{48\pi}\left(1+6|\xi|\right)\sum^5_{i=0}
\triangle _i \mu_i^2ln\mu_i^2+\left\{-\frac{1}{8\pi}
\left(\frac{1}{6}+|\xi|\right)
\left(\frac{1}{6}-\xi\right)\int_{\Sigma} R \right.\nonumber \\ &
+&\left. \int_{\Sigma} \left[\frac{1}{1440\pi }
\left(\sum^2_{a=1}R_{\mu \nu} n^{\mu}_{a}n^{\nu}_{a}-2\sum^2_{a,
b=1}R_{\mu \nu \lambda \rho} n^{\mu}_{a}n^{\nu}_{b}
n^{\lambda}_{a}n^{\rho}_{b}\right) \right]\right\}
\sum^5_{i=0}\triangle _iln\mu_i^2. \label{smu}
\end{eqnarray}
This is the expression of the statistical-mechanical entropy
arising from the nonminimally coupled scalar fields in the
four-dimensional nonextreme stationary axisymmetric black hole. It
is of interest to note that Eq. (\ref{smu}) possesses same form as
the corresponding result of the static black
hole\cite{Solodukhin97}.

\section{thermodynamical entropy  of the nonextreme\\
 stationary axisymmetric black holes} \vspace*{0.5cm}

Mann and Solodukhin \cite{Mann96} showed that an Euclidean
manifold which is obtained by Wick rotation of the Kerr-Newman
geometry has a conical singularity.  By using the conical
singularities method they  obtained the tree-level thermodynamical
entropy $S^{TD}(G_B,c^i_B)$ and its one-loop quantum corrections
$S^{TD}_{div}$ for the Kerr-Newman black hole due to minimally
coupled scalar field. Which are respectively given by
\begin{eqnarray}
S^{TD}(G_B,c^i_B)&=&\frac{A_{\Sigma}}{4G_B}-8\pi \int_{\Sigma}
 \left[\left(c^1_{B}R+\frac{c^2_{B}}{2}\sum^2_{a=1}R_{\mu
\nu}n^\mu _in^\nu _i \right.\right.\nonumber\\& &+\left.\left.
c^3_{B}\sum^2_{a,b=1}R_{\mu \nu \alpha \beta }n^\mu _in^\nu _j
n^\alpha _i n^\beta _j\right)\right], \label{tree}
\end{eqnarray}
\begin{eqnarray} S^{TD}_{div}&=&\frac{A_{\Sigma}}{48\pi \epsilon
^2}+\left\{\frac{1}{144\pi} \int_{\Sigma}
R-\frac{1}{45}\frac{1}{16\pi}\int_{\Sigma} \left(
\sum^2_{a=1}R_{\mu \nu}n^\mu _in^\nu _i-2\sum^2_{a,b=1}R_{\mu \nu
\alpha \beta }n^\mu _in^\nu _j n^\alpha _i n^\beta _j\right)
\right.\nonumber
\\ & &-\left.\frac{1}{90}\frac{1}{16\pi}\int_{\Sigma}
\left(K^aK^a\right)+
\frac{1}{24\pi}\Big(\lambda_1-\frac{\lambda_2}{30}\Big)
\int_{\Sigma}\left(K^aK^a-2tr(K.K)\right) \right\}
ln\frac{L}{\epsilon}, \label{solo0}
\end{eqnarray}
where  $G_B, c^i_{B}, (i=1,2,3)$ represent bare constants
(tree-level), $K^a_{\mu\nu}=-\gamma^{\alpha}_{\mu}
\gamma^{\beta}_{\nu}\nabla_{\alpha}n^a_{\beta}$ is the extrinsic
curvature, and $K^a=g^{\mu \nu }K^a_{\mu \nu}$ is the trace of the
extrinsic curvature \cite{Mann96}. We find that all terms which
relate to the extrinsic curvature in result (\ref{solo0}) are
equal to zero if one insert the Kerr-Newman metric \cite{Mann96}
(or the Einstein-Maxwell dilaton-axion metric \cite{Jing97}) into
them.

Now we proceed to extend the one-loop quantum correction
(\ref{solo0}) to nonminimally coupled scalar fields for the
general nonextreme stationary axisymmetric black hole.

In order to cast the metric (\ref{metric0}) into Mann-Solodukhin's
form \cite{Mann96}, we define a pair of vectors  as follows
\begin{equation}
K=(1, \ \ 0,\ \ 0,\ \ \tilde{\Omega}),\ \ \ \tilde{K}=(F,\ \ \ 0,\
\ \ 0,\ \ \ 1). \label{kk}
\end{equation}
with
\begin{eqnarray} F&=&-\frac{g_{t\varphi}+\tilde{\Omega} g_{\varphi
\varphi}}{g_{tt}+\tilde{\Omega} g_{t\varphi}},
\end{eqnarray}
\begin{eqnarray}
\tilde{\Omega}=\frac{g_{tt}\sqrt{g_{t \varphi}^2-g_{tt}g_{\varphi
\varphi}+g_{tt}g_{\theta \theta}sin^2\theta }}{g_{tt}g_{\varphi
\varphi}-g_{t\varphi}^2-g_{t\varphi}\sqrt{g_{t
\varphi}^2-g_{tt}g_{\varphi \varphi}+g_{tt}g_{\theta
\theta}sin^2\theta }}. \label{omg}
\end{eqnarray}
we find in the region $r_H\leq r< \infty$ that $F=a sin^2\theta $
and $\tilde{\Omega}=\frac{a}{r^2+a^2}$ for the Kerr and the
Kerr-Newman black holes, and $F=a sin^2\theta $ and
$\tilde{\Omega}=\frac{a}{r^2-2dr+a^2}$ for the Einstein-Maxwell
dilaton-axion black hole. In general , on the event horizon,
$\tilde{\Omega}(r_H)=\Omega _H=-\Big(\frac{g_{t\varphi}}
{g_{\varphi \varphi}}\Big)_{r_H}$, where $\Omega _H$ is the
angular velocity of the horizon. We can prove that $K$ and
$\tilde{K}$ are a pair of the Killing vectors on the horizon. What
we need in following is just the properties of the vectors $K$ and
$\tilde{K}$ near the event horizon.

The one forms dual to $K$ and $\tilde{K}$ are respectively given
by
 \begin{equation}
\omega=\left(\frac{1}{1-\tilde{\Omega} F}, \ \ 0,\ \ 0,\ \ \
-\frac{F}{1-\tilde{\Omega} F}\right),\ \ \
\tilde{\omega}=\left(-\frac{\tilde{\Omega}}{1-\tilde{\Omega} F},\
\ \ 0,\ \ \ 0,\ \ \ \frac{1}{1-\tilde{\Omega} F}\right). \label{w}
\end{equation}
Thus, the metric (\ref{metric0}) can be written as
\begin{eqnarray}
ds^2=\Big(g_{tt}+2\tilde{\Omega}g_{t\varphi}+\tilde{\Omega}^2g_{\varphi
\varphi}\Big)\omega ^2+g_{rr}dr^2+g_{\theta \theta}\Big(d\theta
^2+sin^2\theta \tilde{\omega}^2\Big). \label{metricE}
\end{eqnarray}
Euclideanize  the metric by setting $t=i\tau$,
$\tilde{\Omega}=i\hat{\Omega}$, and $F=i\hat{F}$, then the
Euclidean vectors (\ref{kk}) and the corresponding one-forms
(\ref{w}) take the form
\begin{equation}
K=(1, \ \ 0,\ \ 0,\ \ -\hat{\Omega}),\ \ \ \tilde{K}=(\hat{F},\ \
\ 0,\ \ \ 0,\ \ \ 1). \label{kk1}
\end{equation}
\begin{equation}
\omega=\left(\frac{1}{1+\hat{\Omega} \hat{F}}, \ \ 0,\ \ 0,\ \ \
-\frac{\hat{F}}{1+\hat{\Omega} \hat{F}}\right),\ \ \
\tilde{\omega}=\left(\frac{\hat{\Omega}}{1+\hat{\Omega} \hat{F}},\
\ \ 0,\ \ \ 0,\ \ \ \frac{1}{1+\hat{\Omega} \hat{F}}\right).
\label{ww1}
\end{equation}
The metric (\ref{metricE}) becomes
\begin{eqnarray}
ds^2&=&\hat{H}(r, \theta)(r-\hat{r}_H)\Big(d\tau -\hat{F}d\varphi
\Big)^2+\frac{1}{\hat{G}_2(r,\theta)(r-\hat{r}_H)}dr^2\nonumber \\
&+& \hat{g}_{\theta \theta}\Big(d\theta ^2+sin^2\theta
\tilde{\omega}^2\Big), \label{metricE1}
\end{eqnarray}
with
\begin{eqnarray}
\hat{H}(r,\theta)(r-\hat{r}_H)\equiv
\frac{\hat{g}_{tt}+2\hat{\Omega}
\hat{g}_{t\varphi}+\hat{\Omega}^2\hat{g}_{\varphi
\varphi}}{(1+\hat{\Omega}\hat{F})^2}.\nonumber
\end{eqnarray}
It is useful to introduce a new variable
\begin{eqnarray}
(r-\hat{r}_H)=\frac{x^2}{4}, \nonumber
\end{eqnarray}
up to term $o(x^2)$ the metric (\ref{metricE1})  can be expressed
as
\begin{eqnarray}
ds^2=\frac{1}{\hat{G}_2(\hat{r}_H,\theta)}ds^2_{c2}+ds^2_{\Sigma},\label{metricE2}
\end{eqnarray}
with \begin{eqnarray} ds^2_{c2}&=&dx^2+\frac{\hat{H}(\hat{r}_H,
\theta)\hat{G}_2(\hat{r}_H,\theta)x^2}{4}\Big(d\tau
-\hat{F}(\hat{r}_H, \theta)d\varphi \Big)^2,\label{dsc} \\
ds^2_{\Sigma}&=& \hat{g}_{\theta \theta}\Big(d\theta
^2+sin^2\theta \tilde{\omega}^2\Big), \label{dss}
\end{eqnarray}
Introducing new angle coordinate $d\chi=\frac{\beta_H}{\beta
}\sqrt{\frac{4}{\hat{H}\hat{G_2}}}\Big(d\tau-\hat{F}d\varphi
\Big)$, Eq. (\ref{dsc}) reads
\begin{eqnarray} ds^2_{c2\alpha }&=&dx^2+\alpha ^2 x^2 d\chi^2,\label{dsc1}
\end{eqnarray}
From Euclidean metric (\ref{metricE1}) we can define a pair of
vectors orthogonal to the horizon
\begin{eqnarray}
n^{\mu}_1&=&(0, \ \ \ \sqrt{\hat{g}^{rr}},\ \ \  0,\ \ \
0),\nonumber
\\ n^{\mu}_2&=&\left(\frac{1}{\sqrt{\hat{g}_{tt}+2\tilde{\Omega}
\hat{g}_{t\varphi}+\hat{\Omega} ^2\hat{g}_{\varphi \varphi} }},\ \
\ 0, \ \ \ 0,\ \ \
\frac{\hat{\Omega}}{\sqrt{\hat{g}_{tt}+2\hat{\Omega}
\hat{g}_{t\varphi}+\hat{\Omega} ^2\hat{g}_{\varphi \varphi}
}}\right), \label{a1}
\\
n_{\mu}^1&=&\left(0,\ \ \ \frac{1}{\sqrt{\hat{g}^{rr}}}, \ \ \ 0,\
\ \ 0\right), \nonumber
\\
n_{\mu}^2&=&\left(\frac{\sqrt{\hat{g}_{tt}+2\hat{\Omega}
\hat{g}_{t\varphi}+\hat{\Omega} ^2\hat{g}_{\varphi \varphi}
}}{1+\hat{\Omega} \hat{F}},\ \ \ \ 0, \ \ \ 0,\ \ \
-\frac{\hat{F}\sqrt{\hat{g}_{tt}+2\hat{\Omega}
\hat{g}_{t\varphi}+\hat{\Omega} ^2\hat{g}_{\varphi \varphi}
}}{1+\hat{\Omega} \hat{F}}\right). \label{nn}
\end{eqnarray}

It is helpful to note that Eqs. (\ref{dss}), (\ref{metricE2}),
(\ref{dsc1}), (\ref{a1}) and (\ref{nn}) take the similar form as
Eqs. (2.8), (2.9), (3.1), (A1,A2)  and (A3, A4) in
Ref.\cite{Mann96}, respectively. Therefore, the discussions for
the general nonextreme  stationary black hole (\ref{metric0}) is
parallel with  that for the Kerr-Newman black hole\cite{Mann96}.

Following  Solodukhin's \cite{Solodukhin953} and Mann and
Solodukhin's \cite{Mann96} discussions, and employing the results
obtained in Appendix B
\begin{eqnarray}
\Big(K^aK^a\Big)_{r_H}&=&0,\nonumber \\
\Big[tr(K.K)\Big]_{r_H}&=&\Big(K^a_{\mu \nu}K_a^{\mu
\nu}\Big)_{r_H}=0,\label{ka}
\end{eqnarray}
we obtain one-loop quantum correction to the thermodynamical
entropy arising from the nonminimally coupled scalar field for the
general nonextreme  stationary axisymmetric black hole as follows
\begin{eqnarray}
S^{TD}_{div}&=&\frac{A_{\Sigma}}{8\pi\epsilon
^2}\left(\frac{1}{6}-\xi\right)+\left\{\frac{1}{4\pi}
\left(\frac{1}{6}-\xi\right)^2 \int _{\Sigma} R \right.\nonumber
\\ &- & \left.\frac{1}{45}\frac{1}{16\pi} \int _{\Sigma}
 \left(\sum^2_{a=1}R_{\mu \nu}
n^{\mu}_{a}n^{\nu}_{a}-2\sum^2_{a, b=1}R_{\mu \nu \lambda \rho}
n^{\mu}_{a}n^{\nu}_{b} n^{\lambda}_{a}n^{\rho}_{b}\right) \right\}
ln\frac{L}{\epsilon}. \label{smu1}
\end{eqnarray}
The expressions (\ref{ka}) and (\ref{smu1}) has been backed to
real values of the parameters $t$, $\tilde{\Omega}$ and $F$. It is
of interest to note that for the general nonextreme stationary
axisymmetric black hole the contributions of the quadratic
combinations of the extrinsic curvature of the horizon to the
entropy, as for the Kerr-Newman black hole \cite{Mann96},  are
zero when we define a pair of vectors orthogonal to the horizon
$\Sigma$ and dual to one-forms. Therefore, the logarithmically
divergent part of the result (\ref{smu1}) depends only on
projections of the curvature onto subspace normal to the horizon
as the static black hole does.

\section{summary and discussion}

\vspace*{0.5cm}

Since the Pauli-Villars regularization scheme causes a factor
$-\frac{1}{2}$ in second part in the Eq. (\ref{smu}), we know from
above discussions that the statistical-mechanical  entropy
(\ref{smu}) coincides with the
 one-loop quantum correction to thermodynamical entropy (\ref{smu1})
for  $\xi\leq 0$ coupling.

We now seek for relation between the statistical-mechanical
entropy of quantum excitations of the stationary axisymmetric
black hole and its thermodynamical entropy. We first renormalize
the thermodynamical entropy by using standard  approach
\cite{Birrell82} \cite{Solodukhin953} \cite{Demers95}
\cite{Frolov98}. Combing the tree-level entropy (\ref{tree}) with
one-loop correction (\ref{smu1}) one find that the divergence can
be absorbed in the renormalization of the coupling constants
\begin{eqnarray}
\frac{1}{G_{ren}}&=&\frac{1}{G_B}+\frac{1}{2\pi \epsilon
^2}\left(\frac{1}{6}-\xi\right), \nonumber \\
c^1_{ren}&=&c^1_{B}-\frac{1}{32\pi
^2}\left(\frac{1}{6}-\xi\right)^2\ln\frac{L}{\epsilon},\nonumber
\\ c^2_{ren}&=&c^2_{B}+\frac{1}{32\pi
^2}\frac{1}{90}\ln\frac{L}{\epsilon},\nonumber
\\ c^3_{ren}&=&c^3_{B}-\frac{1}{32\pi
^2}\frac{1}{90}\ln\frac{L}{\epsilon}.\label{cc}
\end{eqnarray}
From Eqs.  (\ref{tree}), (\ref{smu1}) and (\ref{cc}) we have
\begin{eqnarray}
& &S^{TD}(G_{ren},c^i_{ren})=S^{TD}(G_{B},c^i_B)+S^{TD}_{div}
\nonumber \\ &=&\frac{A_{\Sigma}}{4G_{ren}}-8\pi \int _{\Sigma}
\left(c^1_{ren}R +\frac{c^2_{ren}}{2}\sum^2_{a=1}R_{\mu \nu}n^\mu
_in^\nu _i+ c^3_{ren}\sum^2_{a,b=1}R_{\mu \nu \alpha \beta }n^\mu
_in^\nu _j n^\alpha _i n^\beta _j\right).\label{ren}
\end{eqnarray}
Since we considered the case that terms quadratic in curvature are
preserved in the renormalized action, the black hole entropy can
be expressed as(see Refs.\cite{Fursaev95}-\cite{Jacobson94})
\begin{eqnarray}
S^{BH}(G_{ren}, c^i_{ren})&=&S^{TD}(G_{ren},c^i_{ren}),\nonumber
\\ S^{BH}(G_{B}, c^i_{B})&=&S^{TD}(G_{B},c^i_{B}),
\end{eqnarray}
and the Bekenstein-Hawking entropy is
\begin{eqnarray}
S^{BH}=\frac{A_{\Sigma}}{4G_{ren}}.
\end{eqnarray}
Noting  $S^{TD}_{div}=S^{SM}$ for $\xi\leq 0$, we obtain for
$\xi\leq 0$ the relation
\begin{eqnarray}
S^{BH}(G_{ren}, c^i_{ren})=S^{BH}(G_B, c^i_{B})+S^{SM},
\label{relationL}
\end{eqnarray}
which agrees the static black hole results shown in Refs.
\cite{Demers95}\cite{Fursaev96}\cite{Larsen96}
\cite{Solodukhin952}\cite{Kabat95}. It is shown that the presence
of the bare pure geometrical contribution $S^{BH}(G_B, c^i_{B})$
evidently excludes the possibility to identify $S^{BH}(G_{ren},
c^i_{ren})$ with $S^{SM}$.

In conclusion, on the one hand by means of the BWM in which the
original Dirichlet condition was replaced by a scattering ansatz
for the field functions at the event horizon and with the
Pauli-Villars regularization scheme, the statistical-mechanical
entropy arising from the nonminimally coupled scalar fields which
rotate with the angular velocity $\Omega_0=\Omega_H$ in the
general four-dimensional nonextreme stationary axisymmetric black
hole space-time is studied. The result can be expressed as either
Eq. (\ref{smuu}) or Eq.  (\ref{smu}).

On the other hand by using the conical singularities method we
extend Mann and Solodukhin's result for the Kerr-Newman black hole
\cite{Mann96} to the general nonextreme stationary axisymmetric
black hole and the nonminimally coupled scalar field.
Nevertheless, we find that the logarithmically part in Eq.
(\ref{smu1}) depends only on projections of the curvature onto
subspace normal to the horizon since the contributions of the
quadratic combinations of the extrinsic curvature of the horizon
are zero when we define a pair of vectors orthogonal to the event
horizon $\Sigma$ and dual to one-forms.

By comparing the statistical-mechanical entropy (\ref{smu}) and
the thermodynamical entropy (\ref{smu1}) we show that, for the
general nonextreme stationary axisymmetric black hole, the
statistical mechanical entropy and the one-loop correction to the
thermodynamic entropy are equivalent for the coupling $\xi\leq 0$.
It is an interesting result that the entropies possesses the same
form as static black hole does\cite{Solodukhin953} if we express
the entropy with the projections of the curvature onto subspace
normal to the horizon.

Combing the tree-level entropy (\ref{tree}) with one-loop
correction  (\ref{smu1}), the divergence can be absorbed in the
renormalization of the gravitational and coupling constants. After
renormalized with standard scheme, a relation between the
statistical-mechanical entropy of quantum excitations of the
nonextreme stationary axisymmetric black hole and its
thermodynamical entropy for the case $\xi \leq 0$ is obtained. The
relation agrees the results for the static black hole and fills in
the gaps mentioned in Ref.\cite{Frolov98}.

\newpage
\appendix

\section{projections of the curvature onto \\
the subspace normal to the horizon surface }

From vectors (\ref{kk}), dual vectors (\ref{w}) and
 metric (\ref{metricE}), we can define a pair of orthonormal vectors
\begin{eqnarray}
n^{\mu}_1&=&(0, \ \ \ \sqrt{g^{rr}},\ \ \  0,\ \ \  0),\nonumber
\\ n^{\mu}_2&=&\left(\frac{1}{\sqrt{-(g_{tt}+2\tilde{\Omega}
g_{t\varphi}+\tilde{\Omega} ^2g_{\varphi \varphi} )}},\ \ \ 0, \ \
\ 0,\ \ \ \frac{\tilde{\Omega}}{\sqrt{-(g_{tt}+2\tilde{\Omega}
g_{t\varphi}+\tilde{\Omega} ^2g_{\varphi \varphi}
)}}\right),\nonumber \\ \label{a11}
\\
n_{\mu}^1&=&\left(0,\ \ \ \frac{1}{\sqrt{g^{rr}}}, \ \ \ 0,\ \ \
0\right), \nonumber
\\
n_{\mu}^2&=&\left(\frac{\sqrt{-(g_{tt}+2\tilde{\Omega}
g_{t\varphi}+\tilde{\Omega} ^2g_{\varphi \varphi}
)}}{1-\tilde{\Omega} F},\ \ \ \ 0, \ \ \ 0,\ \ \
-\frac{F\sqrt{-(g_{tt}+2\tilde{\Omega} g_{t\varphi}+\tilde{\Omega}
^2g_{\varphi \varphi} )}}{1-\tilde{\Omega} F}\right).\nonumber \\
\label{nn1}
\end{eqnarray}
After tediously calculation, the projections of the curvature onto
the subspace normal to the horizon surface
\begin{equation}
R_{aa}=\sum^2_{a=1}R_{\mu \nu} n^{\mu}_{a}n^{\nu}_{a},\ \ \ \
R_{abab}=\sum^2_{a=1}R_{\mu \nu \lambda \rho}
n^{\mu}_{a}n^{\nu}_{b} n^{\lambda}_{a}n^{\rho}_{b}, \label{a2}
\end{equation}
can be expressed by using the metric as
\begin{eqnarray}
R_{aa}(r_H,\theta)&=& \left[\frac{g_{t \varphi}^2}{fg_{\varphi
\varphi}^3}\left(\frac{\partial g_{\varphi \varphi}}{\partial
r}\right)^2 - \frac{2g_{t \varphi}}{fg_{\varphi
\varphi}^2}\frac{\partial g_{\varphi \varphi}}{\partial
r}\frac{\partial g_{t \varphi}}{\partial r} + \frac{1}{fg_{\varphi
\varphi}}\left(\frac{\partial g_{t \varphi}}{\partial r}\right)^2
 \right. \nonumber\\ & & \left.
- \frac{3}{2}\frac{\partial ln f}{\partial r}\frac{\partial
g^{rr}}{\partial r}-\frac{1}{2g_{\theta \theta}}\frac{\partial
g_{\theta \theta}}{\partial r}\frac{\partial g^{rr}}{\partial r} -
\frac{1}{2g_{\varphi \varphi}}\frac{\partial g_{\varphi
\varphi}}{\partial r}\frac{\partial g^{rr}}{\partial r} -
\frac{\partial ^2 g^{rr}}{\partial r^2}\right]_{r_H}. \label{a3}
\\ \nonumber \\ \nonumber
R_{abab}(r_H,\theta)&=& \frac{1}{f}\left[\frac{3g_{t
\varphi}^2}{2g_{\varphi \varphi}^3}\left(\frac{\partial g_{\varphi
\varphi}}{\partial r}\right)^2 - \frac{3g_{t \varphi}}{g_{\varphi
\varphi}^2} \frac{\partial g_{\varphi \varphi}}{\partial
r}\frac{\partial g_{t \varphi}}{\partial r} \right. \nonumber \\ &
&\left. +\frac{3}{2g_{\varphi \varphi}}\left(\frac{\partial g_{t
\varphi}}{\partial r}\right)^2
  -\frac{3}{2}\frac{\partial f}{\partial r}\frac{\partial
g^{rr}}{\partial r} - f\frac{\partial ^2 g^{rr}}{\partial
r^2}\right]_{r_H}.\label{a4}
\end{eqnarray}
Eqs. (\ref{a3}) and (\ref{a4}) yield
\begin{eqnarray}
& &R_{nn}(r_H,\theta)-2R_{mnmn}(r_H,\theta)=\nonumber \\ & &
\left\{\frac{\partial ^2g^{rr}}{\partial r^2}+
\frac{3}{2}\frac{\partial g^{rr}}{\partial r}\frac{\partial \ln
f}{\partial r}-\frac{1}{2}\frac{\partial g^{rr}}{\partial
r}\left(\frac{1} {g_{\theta \theta}}\frac{\partial g_{\theta
\theta}}{\partial r} +\frac{1}{g_{\varphi \varphi}}\frac{\partial
g_{\varphi \varphi}}{\partial r}\right) -\frac{2g_{\varphi
\varphi}}{f} \left[\frac{\partial}{\partial r}\left(\frac{g_{t
\varphi}}{g_{\varphi
\varphi}}\right)\right]^2\right\}_{r_H}.\nonumber \\ \label{a5}
\end{eqnarray}

\section{extrinsic geometry of the horizon}

Using the vectors $n^1$ and $n^2$ defined by Eqs. (\ref{a11}) and
(\ref{nn1}) we can introduce induced metric
\begin{equation}
\gamma _{\mu \nu}=g_{\mu
\nu}-n^1_{\mu}n^1_{\nu}-n^2_{\mu}n^2_{\nu}.
\end{equation}
The nonzero components of the induced metric are
\begin{eqnarray}
\gamma_{tt}&=& \frac{g_{t\varphi} ^2}{g_{\varphi \varphi}}  - f
g^{rr}+ \frac{f Hg^{rr} }{\left( 1 + \frac{\tilde{\Omega}
(g_{t\varphi} + g_{\varphi \varphi}\tilde{\Omega}
)}{g_{t\varphi}^2/g_{\varphi \varphi}- fg^{rr} +
g_{t\varphi}\tilde{\Omega}}\right)^2}, \nonumber
\\ \gamma_{t \varphi}&=&g_{t\varphi}+ \frac{(fg_{\varphi
\varphi}Hg^{rr}
 (g_{t\varphi}+ g_{\varphi \varphi}\tilde{\Omega})(-g_{t\varphi}^2
 + fg_{\varphi \varphi}g^{rr}  - g_{\varphi \varphi}g_{t\varphi}
 \tilde{\Omega}))}{ (-g_{t\varphi}^2 +
fg_{\varphi \varphi}g^{rr}- 2g_{\varphi
\varphi}g_{t\varphi}\tilde{\Omega} - g_{\varphi \varphi
}^2\tilde{\Omega}^2)^2},
 \nonumber\\
\gamma_{\theta \theta}&=&g_{\theta \theta}, \nonumber \\
 \gamma_{\varphi \varphi }&=& g_{\varphi \varphi}+\frac{(fg_{\varphi \varphi}
 ^2Hg^{rr}(g_{t\varphi}+ g_{\varphi \varphi}\tilde{\Omega})^2}
 {(-g_{t\varphi}^2 + fg_{\varphi \varphi}g^{rr}  -2g_{\varphi \varphi}
 g_{t\varphi}\tilde{\Omega}- g_{\varphi
 \varphi}^2\tilde{\Omega}^2)^2}.
\end{eqnarray}
where $H=\left(1+\frac{g_{\varphi \varphi}(\Omega
-\tilde{\Omega})^2}{g_{tt}g_{\varphi
\varphi}-g_{t\varphi}^2}\right)$, and
$\Omega=-\frac{g_{t\varphi}}{g_{\varphi \varphi}}$. With respect
to the normal vectors $n^a_{\mu}, \ \ (a=1, 2)$ we define the
extrinsic curvature $k^a_{\mu
\nu}=-\gamma^{\alpha}_{\mu}\gamma^{\beta}_{\nu}\nabla_{\alpha}n^a_{\beta}$.
The nonzero components of the extrinsic curvature can be expressed
as
\begin{eqnarray}
K^1_{tt}&=& -\frac{\tilde{\Omega}^2 (g_{t \varphi}^2 - f
g_{\varphi \varphi} g^{rr} + g_{\varphi \varphi} g_{t \varphi}
\tilde{\Omega})^2}{2 f^2 g_{\varphi \varphi}^2 H^2 (g^{rr})^{3/2}}
\frac{\partial g_{\varphi \varphi}}{\partial r} \nonumber \\&+&
\frac{\tilde{\Omega} (-g_{t \varphi}^2 + f g_{\varphi \varphi}
g^{rr} - g_{\varphi \varphi} g_{t \varphi} \tilde{\Omega}) (g_{t
\varphi}^2 - f g_{\varphi \varphi} g^{rr} + f g_{\varphi \varphi}
H g^{rr} + g_{\varphi \varphi} g_{t \varphi} \tilde{\Omega})}{ f^2
g_{\varphi \varphi}^2 H^2 (g^{rr})^{3/2}}\frac{\partial g_{t
\varphi}}{\partial r} \nonumber
\\ &+&\frac{(g_{t \varphi}^2 - f g_{\varphi \varphi} g^{rr}
+ f g_{\varphi \varphi} H g^{rr} + g_{\varphi \varphi} g_{t
\varphi} \tilde{\Omega})^2}{2 f^2 g_{\varphi \varphi}^4 H^2
(g^{rr})^{3/2}} \nonumber \\ &\times& \left(g_{\varphi \varphi}^2
g^{rr} \frac{\partial f}{\partial r} + g_{t \varphi}^2
\frac{\partial g_{\varphi \varphi}}{\partial r} - 2 g_{\varphi
\varphi} g_{t \varphi} \frac{\partial g_{t \varphi}}{\partial r} +
f g_{\varphi \varphi}^2 \frac{\partial g^{rr}}{\partial
r}\right),\nonumber\\ \nonumber \\
 K^1_{t\varphi}&=& -\frac{\tilde{\Omega}
(g_{t \varphi}^2 - f g_{\varphi \varphi} g^{rr} + g_{\varphi
\varphi} g_{t \varphi} \tilde{\Omega}) (f H g^{rr} + g_{t \varphi}
\tilde{\Omega} + g_{\varphi \varphi} \tilde{\Omega}^2)}{2 f^2
g_{\varphi \varphi} H^2 (g^{rr})^{3/2}} \frac{\partial g_{\varphi
\varphi}}{\partial r} \nonumber \\ &-& \frac{\tilde{\Omega} (g_{t
\varphi} + g_{\varphi \varphi} \tilde{\Omega}) (g_{t \varphi}^2 -
f g_{\varphi \varphi} g^{rr} + g_{\varphi \varphi} g_{t \varphi}
\tilde{\Omega})} {2 f^2 g_{\varphi \varphi} H^2 (g^{rr})^{3/2}}
\frac{\partial g_{t \varphi}}{\partial r} \nonumber \\ &+&
\frac{(-g_{t \varphi}^2 + f g_{\varphi \varphi} g^{rr} - f
g_{\varphi \varphi} H g^{rr} - g_{\varphi \varphi} g_{t \varphi}
\tilde{\Omega}) (f H g^{rr} + g_{t \varphi} \tilde{\Omega} +
g_{\varphi \varphi} \tilde{\Omega}^2)  }{2 f^2 g_{\varphi \varphi}
H^2 (g^{rr})^{3/2}} \frac{\partial g_{t \varphi}}{\partial r}
\nonumber \\ &+& \frac{(g_{t \varphi} + g_{\varphi \varphi}
\tilde{\Omega}) (g_{t \varphi}^2 - f g_{\varphi \varphi} g^{rr} +
f g_{\varphi \varphi} H g^{rr} + g_{\varphi \varphi} g_{t \varphi}
\tilde{\Omega})}{2 f^2 g_{\varphi \varphi}^3 H^2
(g^{rr})^{3/2}}\nonumber
\\&\times& \left(g_{\varphi \varphi}^2 g^{rr} \frac{\partial
f}{\partial r} + g_{t \varphi}^2 \frac{\partial g_{\varphi
\varphi}}{\partial r} - 2 g_{\varphi \varphi} g_{t \varphi}
\frac{\partial g_{t\varphi}}{\partial r}+f g_{\varphi \varphi}^2
\frac{\partial g^{rr}}{\partial r}\right),\nonumber\\ \nonumber \\
 K^1_{\theta
\theta}&=& -\frac{\sqrt{g^{rr}}}{2}\frac{\partial g_{\theta \theta
}}{\partial r},\nonumber\\ \nonumber \\
 K^1_{\varphi \varphi}&=& -\frac{(f H
g^{rr} + g_{t \varphi} \tilde{\Omega} + g_{\varphi \varphi}
\tilde{\Omega}^2)^2}{ f^2 H^2 (g^{rr})^{3/2}}\left[\frac{1}{2}
\frac{\partial g_{\varphi \varphi}}{\partial r} + \frac{(g_{t
\varphi} + g_{\varphi \varphi} \tilde{\Omega})}{(f H g^{rr} + g_{t
\varphi} \tilde{\Omega} + g_{\varphi \varphi} \tilde{\Omega}^2)}
\frac{\partial g_{t \varphi}}{\partial r}\right] \nonumber \\&&+
(g_{t \varphi} + g_{\varphi \varphi} \tilde{\Omega})^2
\left(g^{rr} \frac{\partial f }{\partial r}+ \frac{g_{t
\varphi}^2}{g_{\varphi \varphi}^2} \frac{\partial g_{\varphi
\varphi}}{\partial r}- \frac{2 g_{t \varphi}}{g_{\varphi \varphi}
\frac{\partial g_{t \varphi}}{\partial r} + f \frac{\partial
g^{rr}}{\partial r}}\right)\frac{1}{2 f^2 H^2
(g^{rr})^{3/2}},\nonumber \\ \label{k1xx}
\end{eqnarray}

\begin{eqnarray}
K^2_{t\theta}&=& \frac{(-f H g^{rr})^{1/2} \tilde{\Omega} (-g_{t
\varphi}^2 + f g_{\varphi \varphi} g^{rr} - g_{\varphi \varphi}
g_{t \varphi} \tilde{\Omega})}{2 f H g^{rr} (-g_{t \varphi}^2 + f
g_{\varphi \varphi} g^{rr} - 2 g_{\varphi \varphi} g_{t \varphi}
\tilde{\Omega} - g_{\varphi \varphi}^2 \tilde{\Omega}^2)}
 (\tilde{\Omega} \frac{\partial g_{\varphi \varphi}}{\partial \theta} +
\frac{\partial g_{t \varphi}}{\partial \theta}) \nonumber\\ &+&
  ((-f H g^{rr})^{1/2} (g_{t \varphi}^2 -
  f g_{\varphi \varphi} g^{rr} + f g_{\varphi \varphi} H g^{rr}
  + g_{\varphi \varphi} g_{t \varphi} \tilde{\Omega})\nonumber\\ &\times&
  \frac{(g_{\varphi \varphi}^2 g^{rr} \frac{\partial f}{\partial \theta}
  + g_{t \varphi}^2
  \frac{\partial g_{\varphi \varphi}}{\partial \theta} -
2 g_{\varphi \varphi} g_{t \varphi} \frac{\partial g_{t
\varphi}}{\partial \theta} - g_{\varphi \varphi}^2 \tilde{\Omega}
\frac{\partial g_{t \varphi}}{\partial \theta} + f g_{\varphi
\varphi}^2 \frac{\partial g^{rr}}{\partial \theta})}{ 2 f
g_{\varphi \varphi}^2 H g^{rr} (-g_{t \varphi}^2 + f g_{\varphi
\varphi} g^{rr} - 2 g_{\varphi \varphi} g_{t \varphi}
\tilde{\Omega} - g_{\varphi \varphi}^2
\tilde{\Omega}^2)},\nonumber \\ \nonumber \\
 K^2_{\varphi
\theta}&=& \frac{g_{\varphi \varphi} (-f H g^{rr})^{1/2} (f H
g^{rr} + g_{t \varphi} \tilde{\Omega} + g_{\varphi \varphi}
\tilde{\Omega}^2)}{2 f H g^{rr} (g_{t \varphi}^2 - f g_{\varphi
\varphi} g^{rr} + 2 g_{\varphi \varphi} g_{t \varphi}
\tilde{\Omega} + g_{\varphi \varphi}^2 \tilde{\Omega}^2)}
\left(\tilde{\Omega} \frac{\partial g_{\varphi \varphi}}{\partial
\theta} + \frac{\partial g_{t \varphi}}{\partial \theta}\right)
\nonumber\\ &&- \left[(-f H g^{rr})^{-1/2} (g_{t \varphi} +
g_{\varphi \varphi} \tilde{\Omega}) \left(g_{\varphi \varphi}^2
g^{rr} \frac{\partial f}{\partial \theta} + g_{t \varphi}^2
\frac{\partial g_{\varphi \varphi}}{\partial \varphi} - 2
g_{\varphi \varphi} g_{t \varphi} \frac{\partial g_{t
\varphi}}{\partial \theta} \right. \right. \nonumber
\\ &&- \left. \left. g_{\varphi \varphi}^2 \tilde{\Omega} \frac{\partial g_{t
\varphi}}{\partial \theta} + f g_{\varphi \varphi}^2
\frac{\partial g^{rr}}{\partial \theta}\right)\right]\frac{1}{2
g_{\varphi \varphi} (-g_{t \varphi}^2 + f g_{\varphi \varphi}
g^{rr} - 2 g_{\varphi \varphi} g_{t \varphi} \tilde{\Omega} -
g_{\varphi \varphi}^2 \tilde{\Omega}^2)}.\nonumber \\ \label{k2xx}
\end{eqnarray}
And the contravariant quantities are given by
\begin{eqnarray}
K_1^{tt}&=&\Big(g_{\varphi \varphi} ^2 g^{rr}  \frac{\partial
f}{\partial r}  -
    2 g_{\varphi \varphi} ^2 H  g^{rr}  \frac{\partial f}{\partial r}  +
    g_{\varphi \varphi} ^2 H ^2 g^{rr}  \frac{\partial f}{\partial r}  +
    g_{t\varphi} ^2 \frac{\partial g_{\varphi \varphi}}{\partial r}  -
        2 g_{t\varphi} ^2 H  \frac{\partial g_{\varphi \varphi}}
        {\partial r} \nonumber \\
  &-&2 g_{\varphi \varphi}  g_{t\varphi}  H  \tilde{\Omega} \frac{\partial
  g_{\varphi \varphi}}{\partial r}  -  g_{\varphi \varphi} ^2
  \tilde{\Omega}^2 \frac{\partial g_{\varphi \varphi}}{\partial r}  -
  2 g_{\varphi \varphi}  g_{t\varphi}  \frac{\partial g_{t\varphi}}{
  \partial r}(1-H) -    2 g_{\varphi \varphi} ^2
  \tilde{\Omega} \frac{\partial g_{t\varphi}}{\partial r}
 \nonumber \\&+&
 2 g_{\varphi \varphi} ^2 H  \tilde{\Omega} \frac{\partial
g_{t\varphi}}{\partial r}  +
    f  g_{\varphi \varphi} ^2 \frac{\partial g^{rr}}{\partial r}(1-H)^2 \Big)/
  (2 f ^2 g_{\varphi \varphi} ^2 H ^2 (g^{rr})^{3/2}),\nonumber\\
\nonumber \\
K_1^{t\varphi}&=&\Big(g_{\varphi \varphi} ^2
g_{t\varphi}  H g^{rr} \frac{\partial f}{\partial r} -
g_{\varphi \varphi} ^2 g_{t\varphi}  H ^2 g^{rr} \frac{\partial
f}{\partial r}  + g_{\varphi \varphi} ^3 g^{rr}  \tilde{\Omega}
\frac{\partial f}{\partial r}(1-H)  +    g_{t\varphi} ^3 H
\frac{\partial g_{\varphi \varphi}}{\partial r} \nonumber \\&-& f
g_{\varphi \varphi}  g_{t\varphi}  H ^2 g^{rr} \frac{\partial
g_{\varphi \varphi}}{\partial r}  + g_{\varphi \varphi}
g_{t\varphi} ^2 \tilde{\Omega}     \frac{\partial g_{\varphi
\varphi}}{\partial r} - f  g_{\varphi \varphi} ^2 H  g^{rr}
\tilde{\Omega} \frac{\partial g_{\varphi \varphi}}{\partial r} -
g_{\varphi \varphi} ^2 g_{t\varphi}  H  \tilde{\Omega}^2
\frac{\partial g_{\varphi \varphi}}{\partial r}  \nonumber \\&-&
g_{\varphi \varphi} ^3 \tilde{\Omega}^3 \frac{\partial g_{\varphi
\varphi}}{\partial r} -    g_{\varphi \varphi} g_{t\varphi} ^2 H
\frac{\partial g_{t\varphi}}{\partial r}  - f g_{\varphi \varphi}
^2 H  g^{rr}  \frac{\partial g_{t\varphi}}{\partial r}  +    f
g_{\varphi \varphi} ^2 H ^2 g^{rr}  \frac{\partial
g_{t\varphi}}{\partial r}  \nonumber\\&-&    2 g_{\varphi \varphi}
^2 g_{t\varphi}  \tilde{\Omega} \frac{\partial
g_{t\varphi}}{\partial r}  -   2 g_{\varphi \varphi} ^3
\tilde{\Omega}^2 \frac{\partial g_{t\varphi}}{\partial r} +
g_{\varphi \varphi} ^3 H \tilde{\Omega}^2 \frac{\partial
g_{t\varphi}}{\partial r}  + f g_{\varphi \varphi} ^2 g_{t\varphi}
H  \frac{\partial g^{rr}}{\partial r}(1-H) \nonumber\\ &+&   f
g_{\varphi \varphi} ^3 \tilde{\Omega} \frac{\partial
g^{rr}}{\partial r} -    f g_{\varphi \varphi} ^3 H \tilde{\Omega}
\frac{\partial g^{rr}}{\partial r} \Big)/
  (2 f ^2 g_{\varphi \varphi} ^3 H ^2 (g^{rr})^{3/2}), \nonumber\\
\nonumber \\ K_1^{\theta \theta}&=&-\Big((g^{rr})^{1/2}
\frac{\partial g_{\theta \theta }}{\partial r}\Big)/(2 g_{\theta
\theta} ^2), \nonumber \\ \nonumber \\
K_1^{\varphi\varphi}&=&\Big(g_{\varphi \varphi} g_{t\varphi} ^2 H
^2 g^{rr} \frac{\partial f}{\partial r}  +    2 g_{\varphi
\varphi} ^2 g_{t\varphi}  H  g^{rr}  \tilde{\Omega} \frac{\partial
f}{\partial r}  +    g_{\varphi \varphi} ^3 g^{rr}
\tilde{\Omega}^2 \frac{\partial f}{\partial r}  +    2 f
g_{t\varphi} ^2 H ^2 g^{rr}  \frac{\partial g_{\varphi
\varphi}}{\partial r}  \nonumber\\ &-&    f ^2 g_{\varphi
\varphi}H ^2 (g^{rr})^2 \frac{\partial g_{\varphi
\varphi}}{\partial r}  + 2 g_{t\varphi} ^3 H  \tilde{\Omega}
\frac{\partial g_{\varphi \varphi}}{\partial r}  +    g_{\varphi
\varphi}  g_{t\varphi} ^2 \tilde{\Omega}^2 \frac{\partial
g_{\varphi \varphi}}{\partial r} + 2 g_{\varphi \varphi}
g_{t\varphi} ^2 H  \tilde{\Omega}^2 \frac{\partial g_{\varphi
\varphi}}{\partial r}  \nonumber\\&-& 2 f g_{\varphi \varphi} ^2 H
g^{rr}  \tilde{\Omega}^2 \frac{\partial g_{\varphi
\varphi}}{\partial r}  -    g_{\varphi \varphi} ^3
\tilde{\Omega}^4 \frac{\partial g_{\varphi \varphi}}{\partial r} -
2 f  g_{\varphi \varphi}  g_{t\varphi}  H ^2 g^{rr} \frac{\partial
g_{t\varphi}}{\partial r}  - 2 g_{\varphi \varphi} g_{t\varphi} ^2
H  \tilde{\Omega}     \frac{\partial g_{t\varphi}}{\partial r}
\nonumber\\&-& 2 f  g_{\varphi \varphi} ^2 H g^{rr} \tilde{\Omega}
\frac{\partial g_{t\varphi}}{\partial r}  - 2 g_{\varphi \varphi}
^2 g_{t\varphi} \tilde{\Omega}^2     \frac{\partial
g_{t\varphi}}{\partial r}  - 2 g_{\varphi \varphi} ^2 g_{t\varphi}
H  \tilde{\Omega}^2 \frac{\partial g_{t\varphi}}{\partial r}  - 2
g_{\varphi \varphi} ^3 \tilde{\Omega}^3 \frac{\partial
g_{t\varphi}}{\partial r} \nonumber\\ &+&    f  g_{\varphi
\varphi}  g_{t\varphi} ^2 H ^2 \frac{\partial g^{rr}}{\partial r}
+    2 f  g_{\varphi \varphi} ^2 g_{t\varphi}  H  \tilde{\Omega}
\frac{\partial g^{rr}}{\partial r}  +    f  g_{\varphi \varphi} ^3
\tilde{\Omega}^2 \frac{\partial g^{rr}}{\partial r} \Big)/  (2 f
^2 g_{\varphi \varphi} ^3 H ^2 (g^{rr})^{3/2}),\nonumber \\
\label{k1ss} \nonumber \\
 K_2^{t
\theta}&=&\Big[(-f  H  g^{rr} )^{1/2}    \Big(2 g_{\varphi
\varphi} ^2 g_{t\varphi} ^2 g^{rr}  \frac{\partial f}{\partial
\theta}  - 2 g_{\varphi \varphi} ^2 g_{t\varphi} ^2 H  g^{rr}
\frac{\partial f}{\partial \theta}  -      f  g_{\varphi \varphi}
^3 (g^{rr})^2 \frac{\partial f}{\partial \theta}  \nonumber \\ &+&
f g_{\varphi \varphi} ^3 H  (g^{rr})^2 \frac{\partial f}{\partial
\theta}  +      2 g_{\varphi \varphi} ^3 g_{t\varphi}  g^{rr}
\tilde{\Omega} \frac{\partial f}{\partial \theta}  -      2
g_{\varphi \varphi} ^3 g_{t\varphi}  H  g^{rr}  \tilde{\Omega}
\frac{\partial f}{\partial \theta}  +      2 g_{t\varphi} ^4
\frac{\partial g_{\varphi \varphi}}{\partial \theta} \nonumber\\
&-& f  g_{\varphi \varphi}  g_{t\varphi} ^2 g^{rr} \frac{\partial
g_{\varphi \varphi}}{\partial \theta}  -      f g_{\varphi
\varphi}  g_{t\varphi} ^2 H  g^{rr} \frac{\partial g_{\varphi
\varphi}}{\partial \theta}  + 4 g_{\varphi \varphi} g_{t\varphi}
^3 \tilde{\Omega} \frac{\partial g_{\varphi \varphi}}{\partial
\theta}  \nonumber\\ &-& (2+H) f  g_{\varphi \varphi} ^2
g_{t\varphi} g^{rr} \tilde{\Omega} \frac{\partial g_{\varphi
\varphi}}{\partial \theta}   + 2 g_{\varphi \varphi} ^2
g_{t\varphi} ^2 \tilde{\Omega}^2 \frac{\partial g_{\varphi
\varphi}}{\partial \theta} - f g_{\varphi \varphi} ^3 g^{rr}
\tilde{\Omega}^2 \frac{\partial g_{\varphi \varphi}}{\partial
\theta}   \nonumber\\ &-& 2 g_{\varphi \varphi}  g_{t\varphi} ^3
\frac{\partial g_{t\varphi}}{\partial \theta}  +      f g_{\varphi
\varphi} ^2 g_{t\varphi}H  g^{rr} \frac{\partial
g_{t\varphi}}{\partial \theta}  - 4 g_{\varphi \varphi} ^2
g_{t\varphi} ^2 \tilde{\Omega} \frac{\partial
g_{t\varphi}}{\partial \theta} + f  g_{\varphi \varphi} ^3 H
g^{rr}  \tilde{\Omega} \frac{\partial g_{t\varphi}}{\partial
\theta}  \nonumber\\&-& 2 g_{\varphi \varphi} ^3 g_{t\varphi}
\tilde{\Omega}^2 \frac{\partial g_{t\varphi}}{\partial \theta}  +
2 f  g_{\varphi \varphi} ^2 g_{t\varphi} ^2       \frac{\partial
g^{rr}}{\partial \theta}  - 2 f  g_{\varphi \varphi} ^2
g_{t\varphi} ^2 H \frac{\partial g^{rr}}{\partial \theta}  - f ^2
g_{\varphi \varphi} ^3 g^{rr} \frac{\partial g^{rr}}{\partial
\theta} \nonumber\\ &+& f ^2 g_{\varphi \varphi} ^3 H  g^{rr}
\frac{\partial g^{rr}}{\partial \theta}  + 2(1-H) f g_{\varphi
\varphi} ^3 g_{t\varphi} \tilde{\Omega} \frac{\partial
g^{rr}}{\partial \theta}  \Big)\Big]/ \Big(2 f ^2 g_{\theta
\theta} g_{\varphi \varphi} ^2 H (g^{rr})^2 (g_{t\varphi} ^2
\nonumber \\&-& f g_{\varphi \varphi} g^{rr}  + g_{\varphi
\varphi} g_{t\varphi} \tilde{\Omega} + g_{\varphi \varphi}
g_{t\varphi} \tilde{\Omega} + g_{\varphi \varphi} ^2
\tilde{\Omega} \tilde{\Omega} )\Big), \nonumber\\ \nonumber \\
 K_2^{\varphi
\theta}&=&\Big[(-f H g^{rr} )^{1/2} \Big(-2 g_{\varphi \varphi}
g_{t\varphi} ^3 H g^{rr} \frac{\partial f}{\partial \theta}  + f
g_{\varphi \varphi} ^2 g_{t\varphi}  H (g^{rr})^2 \frac{\partial
f}{\partial \theta} - 2 g_{\varphi \varphi} ^2 g_{t\varphi} ^2
g^{rr}  \tilde{\Omega} \frac{\partial f}{\partial \theta}
\nonumber\\&-& 2 g_{\varphi \varphi} ^2 g_{t\varphi} ^2 H g^{rr}
\tilde{\Omega} \frac{\partial f}{\partial \theta}  +      f
g_{\varphi \varphi} ^3 (g^{rr})^2 \tilde{\Omega} \frac{\partial
f}{\partial \theta}  - 2 g_{\varphi \varphi} ^3 g_{t\varphi}
g^{rr}  \tilde{\Omega}^2 \frac{\partial f}{\partial \theta}  - 3 f
g_{t\varphi} ^3 H g^{rr} \frac{\partial g_{\varphi
\varphi}}{\partial \theta}  \nonumber\\&+& 2 f ^2 g_{\varphi
\varphi} g_{t\varphi}  H  (g^{rr})^2 \frac{\partial g_{\varphi
\varphi}}{\partial \theta}  - 2 g_{t\varphi} ^4 \tilde{\Omega}
\frac{\partial g_{\varphi \varphi}}{\partial \theta}  + f(1-3H)
g_{\varphi \varphi} g_{t\varphi} ^2 g^{rr} \tilde{\Omega}
\frac{\partial g_{\varphi \varphi}}{\partial \theta}  \nonumber\\
&+& f ^2 g_{\varphi \varphi} ^2 H  (g^{rr})^2 \tilde{\Omega}
\frac{\partial g_{\varphi \varphi}}{\partial \theta}  -      4
g_{\varphi \varphi} g_{t\varphi} ^3 \tilde{\Omega}^2
\frac{\partial g_{\varphi \varphi}}{\partial \theta}  +      2 f
g_{\varphi \varphi} ^2 g_{t\varphi}  g^{rr} \tilde{\Omega}^2
\frac{\partial g_{\varphi \varphi}}{\partial \theta}
\nonumber\\&+& f g_{\varphi \varphi} ^3 g^{rr} \tilde{\Omega}^3
\frac{\partial g_{\varphi \varphi}}{\partial \theta}  + 3 f
g_{\varphi \varphi} g_{t\varphi} ^2 H       g^{rr} \frac{\partial
g_{t\varphi}}{\partial \theta}  -      f ^2 g_{\varphi \varphi} ^2
H  (g^{rr})^2 \frac{\partial g_{t\varphi}}{\partial \theta}  + 2
g_{\varphi \varphi} g_{t\varphi} ^3 \tilde{\Omega} \frac{\partial
g_{t\varphi}}{\partial \theta}  \nonumber\\&+&      3 f g_{\varphi
\varphi} ^2 g_{t\varphi}  H  g^{rr}  \tilde{\Omega} \frac{\partial
g_{t\varphi}}{\partial \theta}  + 4 g_{\varphi \varphi} ^2
g_{t\varphi} ^2 \tilde{\Omega}^2       \frac{\partial
g_{t\varphi}}{\partial \theta}  + 2 g_{\varphi \varphi} ^3
g_{t\varphi}  \tilde{\Omega}^3       \frac{\partial
g_{t\varphi}}{\partial \theta}  - 2 f  g_{\varphi \varphi}
g_{t\varphi} ^3 H       \frac{\partial g^{rr}}{\partial \theta}
\nonumber\\&+& f ^2 g_{\varphi \varphi} ^2 g_{t\varphi}  H g^{rr}
\frac{\partial g^{rr}}{\partial \theta}  -      2 f g_{\varphi
\varphi} ^2 g_{t\varphi} ^2 \tilde{\Omega} \frac{\partial
g^{rr}}{\partial \theta}  -      2 f  g_{\varphi \varphi} ^2
g_{t\varphi} ^2 H  \tilde{\Omega}       \frac{\partial
g^{rr}}{\partial \theta}  + f ^2 g_{\varphi \varphi} ^3 g^{rr}
\tilde{\Omega}       \frac{\partial g^{rr}}{\partial \theta}
\nonumber\\&-& 2 f  g_{\varphi \varphi} ^3 g_{t\varphi}
\tilde{\Omega}^2 \frac{\partial g^{rr}}{\partial \theta}
 - 2
g_{\varphi \varphi} ^2 g_{t\varphi} ^2 \tilde{\Omega}^3
\frac{\partial g_{\varphi \varphi}}{\partial \theta}\Big)\Big]/
\Big(2 f ^2 g_{\theta \theta}  g_{\varphi \varphi} ^2 H (g^{rr})^2
(-g_{t\varphi} ^2 -      g_{\varphi \varphi} ^2 \tilde{\Omega}^2
 \nonumber\\&-& g_{\varphi \varphi} g_{t\varphi}
\tilde{\Omega}  + f  g_{\varphi \varphi}  g^{rr}
\Big).\label{k2ss}
\end{eqnarray}
The trace of the extrinsic curvature $k^a=g^{\mu \nu}k^a_{\mu \nu
}$ are
\begin{eqnarray}
K^1&=&\frac{1}{2f^2g_{\theta \theta}g_{\varphi
\varphi}^3H^2(g^{rr})^{3/2}}\Big(g_{\theta \theta}g_{\varphi
\varphi}^2g_{t\varphi}^2g^{rr}\frac{\partial  f}{\partial r} -
f(1-H)^2g_{\theta \theta}g_{\varphi
\varphi}^3(g^{rr})^2\frac{\partial f}{\partial r} \nonumber
\\&+& 2g_{\theta
\theta}g_{\varphi
\varphi}^3g_{t\varphi}g^{rr}\tilde{\Omega}\frac{\partial
f}{\partial r} + g_{\theta \theta}g_{\varphi
\varphi}^4g^{rr}\tilde{\Omega}^2\frac{\partial  f}{\partial r} -
 f^2g_{\varphi \varphi}^3H^2(g^{rr})^2\frac{\partial  g_{\theta \theta}}{\partial
r} \nonumber
\\&+& g_{\theta \theta}g_{t\varphi}^4\frac{\partial  g_{\varphi
\varphi}}{\partial r} - f(1-2H)g_{\theta \theta}g_{\varphi
\varphi}g_{t\varphi}^2g^{rr}\frac{\partial  g_{\varphi
\varphi}}{\partial r}   - f^2g_{\theta \theta}g_{\varphi
\varphi}^2H^2(g^{rr})^2\frac{\partial  g_{\varphi
\varphi}}{\partial r} \nonumber\\&+& 2g_{\theta \theta}g_{\varphi
\varphi}g_{t\varphi}^3\tilde{\Omega}\frac{\partial  g_{\varphi
\varphi}}{\partial r} + f(1-2H)g_{\theta \theta}g_{\varphi
\varphi}^3g^{rr}\tilde{\Omega}^2\frac{\partial  g_{\varphi
\varphi}}{\partial r} - 2g_{\theta \theta}g_{\varphi
\varphi}^3g_{t\varphi}\tilde{\Omega}^3\frac{\partial  g_{\varphi
\varphi}}{\partial r} \nonumber \\&-& g_{\theta \theta}g_{\varphi
\varphi}^4\tilde{\Omega}^4\frac{\partial  g_{\varphi
\varphi}}{\partial r} - 2g_{\theta \theta}g_{\varphi
\varphi}g_{t\varphi}^3\frac{\partial  g_{t\varphi}}{\partial r} +
2f(1-2H)g_{\theta \theta}g_{\varphi
\varphi}^2g_{t\varphi}g^{rr}\frac{\partial  g_{t\varphi}}{\partial
r} \nonumber
\\&-& 6g_{\theta \theta}g_{\varphi
\varphi}^2g_{t\varphi}^2\tilde{\Omega}\frac{\partial
g_{t\varphi}}{\partial r} + 2f(1-2H)g_{\theta \theta}g_{\varphi
\varphi}^3g^{rr}\tilde{\Omega}\frac{\partial
g_{t\varphi}}{\partial r}  - 6g_{\theta \theta}g_{\varphi
\varphi}^3g_{t\varphi}\tilde{\Omega}^2\frac{\partial
g_{t\varphi}}{\partial r} \nonumber \\&-& 2g_{\theta
\theta}g_{\varphi \varphi}^4\tilde{\Omega}^3\frac{\partial
g_{t\varphi}}{\partial r} + fg_{\theta \theta}g_{\varphi
\varphi}^2g_{t\varphi}^2\frac{\partial g^{rr}}{\partial r} -
f^2g_{\theta \theta}g_{\varphi \varphi}^3g^{rr}\frac{\partial
g^{rr}}{\partial r} + 2f^2g_{\theta \theta}g_{\varphi
\varphi}^3Hg^{rr}\frac{\partial g^{rr}}{\partial r} \nonumber
\\&-& f^2g_{\theta \theta}g_{\varphi
\varphi}^3H^2g^{rr}\frac{\partial g^{rr}}{\partial r} +
2fg_{\theta \theta}g_{\varphi
\varphi}^3g_{t\varphi}\tilde{\Omega}\frac{\partial
g^{rr}}{\partial r} + fg_{\theta \theta}g_{\varphi
\varphi}^4\tilde{\Omega}^2\frac{\partial g^{rr}}{\partial
r}\Big),\label{kkk2} \nonumber \\
 K^2&=&0.
\end{eqnarray}
All quantities listed by Eqs. (\ref{k1xx}), (\ref{k2xx}),
(\ref{k1ss}), (\ref{k2ss}), and (\ref{kkk2}) are equal to zero (
they are proportional to $\sqrt{g^{rr}}$ near the event horizon)
on the event horizon. Consequently, on the event horizon, we
obtain
\begin{eqnarray}
 K^aK^a&=&0, \nonumber \\
 tr(K.K)&=&K^a_{\mu \nu}K_a^{\mu \nu}=0.
\end{eqnarray}
which show that the quadratic combinations of the extrinsic
curvature are zero on the event horizon  when we use the vectors
(\ref{nn1}). The calculation for the Kerr-Newman and
Einstein-Maxwell dilaton-axion black holes in Refs. \cite{Mann96}
\cite{Jing96} supported the result.


\begin{references}

\bibitem{Frolov98}V. P. Frolov and D. V. Fursaev,
Class. \ Quantum \ Grav.  {\bf 15}, 2041 (1998).

\bibitem{Solodukhin951}S. N. Solodukhin,
Phys.\ Rev.\ D {\bf 51}, 618 (1995).

\bibitem{Fursaev941}D. V. Fursaev,
Class. Quantum \ Grav.   {\bf 11}, 1431 (1994 ).

\bibitem{Cognola94}G. Cognola, K. Kirsten, and L. Vanzo,
Phys. \ Rev. \ D  {\bf 49}, 1029 (1994 ).

\bibitem{Solodukhin952}S. N. Solodukhin,
Phys.\ Rev.\ D {\bf 51}, 609 (1995).

\bibitem{Solodukhin953}S. N. Solodukhin,
Phys.\ Rev.\ D {\bf 52}, 7046 (1995).

\bibitem{Barvinsky96}A. O. Barvinsky and S. N. Solodukhin,
Nucl. \ Phys.\ B {\bf 479}, 305 (1996).

\bibitem{Fursaev94}D. V. Fursaev,
Phys. \ Lett. \ B  {\bf 334}, 53 (1994 ).

\bibitem{Fursaev951}D. V. Fursaev, Mod. \ Phys. \ Lett.
A{\bf 10}, 649 (1995).

\bibitem{Fursaev96}D. V. Fursaev and S. N. Solodukhin,
Phys. \ Lett.  B {\bf 365}, 51 (1996).

\bibitem{Mann96}R. B. Mann and S. N. Solodukhin,
Phys. \ Rev.  D {\bf 54}, 3932 (1996).


\bibitem{Hooft85}G. 't. Hooft,
Nucl. \ Phys.\ B {\bf 256}, 727 (1985).

\bibitem{Solodukhin97}S. N. Solodukhin,
Phys.\ Rev.\ D {\bf 56}, 4968 (1997).

\bibitem{Demers95}J. G. Demers, R. Lafrance, and R.C. Myers,
Phys.\ Rev.\ D {\bf 52}, 2245 (1995).

\bibitem{Ghosh94}A. Ghosh and  P. Mitra,
Phys.\ Rev.\ Lett. {\bf 73}, 2521 (1994).

\bibitem{Lee96}M.H. Lee and J. K. Kim,
Phys.\ Lett.\ A {\bf 212}, 323 (1996).

\bibitem{Mann92}R. B. Mann, L. Tarasov, and A. Zelnikov,
Class. \ Quantum \ Grav.  {\bf 9}, 1487 (1992).

\bibitem{Romeo96}A. Romeo,
Class. \ Quantum \ Grav.  {\bf 13}, 2797 (1996).

\bibitem{Kim97}S. P. Kim {\rm et al}
Phys.  \ Rev. D {\bf 55}, 2159 (1997).

\bibitem{Cognola98}G. Cognola and P. Lecca,
Phys.  \ Rev. D {\bf 57}, 2159 (1998).

\bibitem{Lee961}M.H. Lee and J. K. Kim,
Phys.\ Rev.\ D {\bf 54}, 3904 (1996).

\bibitem{Lee96b}M.H. Lee, H. C. Kim, and J. K. Kim,
Phys.\ Lett.\ B {\bf 388}, 487 (1996).

\bibitem{Ho96}J. Ho, W. T. Kim and Y. J. Park,
"Entropy in the Kerr-Newman black hole". gr-qc/9704032, 1997.

\bibitem{Kim971}S. W. Kim, W. T. Kim, Y. J. Park, and H. Shin,
Phys.\ Lett.\ B {\bf 392}, 311 (1997).

\bibitem{Belgiorno96}F. Belgiorno and S. Liberati, Phys.\ Rev.\ D
{\bf 53}, 3172 (1996).

\bibitem{Jing98}Jiliang Jing,
Int. J. Theor. Phys.  {\bf 37} 1441 (1998).

\bibitem{Jing97}Jiliang Jing,
Chin. \ Phys. \ Lett. {\bf 14}, 495 (1997).

\bibitem{Solodukhin96}S. N. Solodukhin,
Phys.\ Rev.\ D {\bf 54}, 3900 (1996).


\bibitem{Frolov96b}V. P. Frolov, D. V. Fursaev, and A. I. Zelnikov,
Phys.\ Lett.\ B {\bf 382}, 220 (1996).

\bibitem{Frolov96d}V. P. Frolov, D. V. Fursaev, and A. I. Zelnikov,
Phys.\ Rev.\ D {\bf 54}, 2711 (1996).

\bibitem{Fursaev97}D. V. Fursaev, Euclidean and Canonical
Formulations of Statistical Mechanics in the Presence of Killing
Horizon, hep-th/9709213.

\bibitem{Jing99}Jiliang Jing and Mu-Lin Yan, gr-qc/9904001;
Phys. \ Rev. D{\bf }60, (1999) 084015.

\bibitem{Jing96}Jiliang Jing,
Nucl. \ Phys.B {\bf 476}, 548 (1996).

\bibitem{Birrell82} N. D. Birrell and P.C.W.Devies, Quantum Fields in
Curved Space (Cambridge University Press, Cambridge,
England,1982).

\bibitem{Fursaev95}D. V. Fursaev and S. N. Solodukhin,
Phys.\ Rev.\ D {\bf 52}, 2133 (1995).

\bibitem{Wald}R. M. Wald,
Phys.\ Rev.\ D {\bf 48}, R3427 (1993).

\bibitem{Iyer94}V. Iyer and R. M. Wald,
Phys.\ Rev.\ D {\bf 50}, 846 (1994).

\bibitem{Jacobson94}T. A. Jacobson, G. Kang, and R. C. Myers,
Phys.\ Rev.\ D {\bf 49}, 6587 (1994).

\bibitem{Larsen96}F. Larsen and F. Wilczek,
Nucl. \ Phys.\ B {\bf 458}, 249 (1996).

\bibitem{Kabat95}D. Kabat, Nucl. Phys.
Nucl. \ Phys.\ B {\bf 453}, 281 (1995).
\end{references}
\end{document}